# Epitaxial Growth of Pentacene on Alkali Halide Surfaces Studied by Kelvin Probe Force Microscopy

Julia L. Neff[†], Peter Milde[‡], Carmen Pérez León[†], Matthew D. Kundrat[§], Christoph R. Jacob[§], Lukas Eng[‡], Regina Hoffmann-Vogel[*,†]

[†]Physikalisches Institut and DFG-Center for Functional Nanostructures, Karlsruhe Institute of Technology, Wolfgang-Gaede-Str. 1, 76131 Karlsruhe, Germany

[‡]Institut für Angewandte Photophysik, Technical University of Dresden, George-Baehr-Str. 1, 01069 Dresden, Germany

[§]Institute of Physical Chemistry and DFG-Center for Functional Nanostructures, Karlsruhe Institute of Technology, Wolfgang-Gaede-Str. 1a, 76131 Karlsruhe, Germany

## Abstract

In the field of molecular electronics thin films of molecules adsorbed on insulating surfaces are used as the functional building blocks of electronic devices. A control of the structural and electronic properties of the thin films is required for a reliable operating mode of such devices. Here, noncontact atomic force and Kelvin probe force microscopies have been used to investigate the growth and electronic properties of pentacene on KBr(001) and KCl(001) surfaces. Mainly molecular islands of upright standing pentacene are formed, whereas a new phase of tilted molecules appear near step edges on some KBr samples. Local contact potential differences (LCPD) have been studied with both Kelvin experiments and density-functional theory calculations. Large LCPD are found between the substrate and the differently oriented molecules, which may be explained by a partial charge transfer from the pentacene to the surface. The monitoring of the changes of the pentacene islands during dewetting shows that multilayers build up at the expense of monolayers. Moreover, in the Kelvin images, previously unknown line defects appear, which unveil the epitaxial growth of pentacene crystals.

## Introduction

Molecular electronics offers an alternative to conventional silicon electronics where individual or groups of molecules can be used as the functional building blocks of electronic devices. For a reliable functioning of such devices, it is needed, on the one hand, the decoupling of the electronic structure of the molecules from that of the substrate.[1] This can be achieved by using insulating surfaces as substrates, e.g. alkali halides.[2,3] On the other hand, a control of the structural and electronic properties of the molecular layer[4–6] is also required. Of special interest is the initial stage of formation of thin films and their stability, i.e. their dewetting properties.[7] The main techniques used for characterizing molecules on surfaces, such as photoemission and scanning tunneling microscopy, rely on the interaction of electrons with the surfaces, thus, most studies have been limited to conductive surfaces. The development of scanning force microscopy permits the identification of the structural properties of molecules on insu-lators.[8] Moreover, Kelvin probe force microscopy (KPFM) has the capability to detect surface potential variations down to nanometer resolution.[9,10] In KPFM, a scanning force microscope is used to measure the electrostatic forces on the sample surface. For metals, these forces originate from work function differences between the tip and the sample. Applying an appropriate voltage to the sample or to the tip, the electrostatic forces can be minimized and the contact potential differences (CPD) determined. If the sample is covered by a thin overlayer of another material, the work function can change, e.g., due to electron transfer and structural relaxation at the interface. Only recently, KPFM has started to be applied to insulating materials[11] and molecules deposited on bulk insulators.[12–14] On semiconducting or insulating materials, not only work function and ionization energy differences can cause such electrostatic forces but also localized charge, e.g., from charge transfer or interface dipoles. Electrochemical equilibrium is reached when the Fermi level between the tip, the sample and its back electrode are aligned. In the case of wide band gap insulators, e.g. alkali halide crystals, this equilibrium is reached only after long times such that the bulk Fermi level may not be well-defined.[15,16] The absolute values of the CPD can therefore vary from measurement to measurement, hence, it is more appropriate to focus on the variation of the contact potential differences along the surface (LCPD) rather than on its absolute value.[16]

[*]To whom correspondence should be addressed

In this work, we have investigated the structure and electrostatic landscape of pentacene islands on KBr(001) and KCl(001) surfaces by dynamic force microscopy and Kelvin probe force microscopy. Pentacene, among other organic molecules, has shown to be a promising p-type organic semiconductor which can be used to produce organic thin film transistors.[17] Our results reveal that besides the well-known phases of upright standing molecules, a phase of tilted pentacene is formed on some KBr samples. We also have found that differently oriented phases of the molecules cause different interface dipole or charge densities on the surface. DFT studies allow us to tentatively attribute these electronic effects to partial charge transfer between the molecules and the insulator surface. The dewetting of the pentacene results in a change of the morphology of the islands and mutilayers grow at the expense of monolayer islands. KPFM images reflect characteristic line defects on the islands which are related to the orientational growth of the islands. In particular they are oriented parallel to the point-on-line epitaxy directions. Such lines are observed on mono- and multilayer islands, but when they appear on multilayer islands, the lines always run along their middle axis, confirming their relation with the dewetting and crystallization process.

# Results and discussion

Figures 1 and 2 show the KBr and KCl (001) surfaces after molecular deposition at room temperature. The majority of the islands display an apparent height of $1.65 \pm 0.10$ nm; see linecuts in Figure 1(e) and 2(c). This length corresponds approximately to the length of a molecule, pointing to an upright standing configuration of pentacene (type 1 islands). The height measurements were cross-calibrated with the apparent height of substrate steps (0.33 nm for KBr(001) and 0.31 nm for KCl(001)), and determined from measurements where the electrostatic forces were compensated by using KPFM. The islands grow across substrate step edges by shifting the molecular layer vertically without additional deformations detectable at this scale.

Previous works devoted to the growth of pentacene on alkali halides with X-ray diffraction, high-resolution electron diffraction and AFM observed similar islands.[18–21] Our group has previously published high-resolution NC-AFM images of pentacene on KCl(001).[21] Layers and islands of nearly upright standing molecules on the KBr(001)[18] and KCl(001) surfaces[18–20] have been reported in mainly two phases: the thin film[22,23] and the bulk phases.[24–26] While molecules in the thin film phase are oriented nearly perpendicular to the substrate, in the bulk phase they show a tilting angle of approximately 75°, thus displaying a slightly smaller height.[27,28]

Together with the islands of upright standing molecules, in some KBr samples a second type of island with a height of $0.45 \pm 0.05$ nm has been found (type 2), e.g. Figure 1(c) and (e). Owing to their reduced height, we assume that the pentacene molecules in these type 2 islands are arranged in a flat-lying or tilted fashion. On single crystalline metal surfaces, e.g. HOPG, Ag(111), Au(111) or Cu(110), phases of flat-lying molecules are well known.[29–33] However, on HOPG this flat configuration is present only for submonolayer growth.[32] Upon further deposition of molecules, when an overlayer start to grow on top of the flat-lying molecules, those on the first layer rotate around their long molecular axis, changing their flat configuration into tilted fashion.[32] Such a structural modification is unusual and was explained as a consequence of the weak interaction of the molecules with the graphite.[32] On KBr(001) and KCl(001) no strong interaction of the molecules with the surface is expected. Thus, we believe that the adsorbed molecules are arranged in a tilted fashion, which can be stabilized by molecule-molecule interactions, rather than in a flat configuration. Probably pentacene forms rows of parallel molecules due to $\pi-\pi$ stacking, as suggested in the model in Figure 1(e). This configuration is consistent with the measured apparent height of 0.45 nm and a tilting angle of approximately 30°, which agrees with the one found in the crystal structure of the bulk and thin film phases.[23–25] Type 2 islands do not grow over step edges but are confined on a terrace. They appear to grow aligned along the step edges. Presumably, they grow from the step edges of the terraces which are not exactly oriented along the <100> directions, which are the preferred orientations of the step edges on the (001) surfaces. Hence, this type of island is only observed for the case in which the substrate has a large amount of higher indexed surface steps. This seems to be plausible, since it has been reported that molecular layers form special structures by confinement through substrate step edges.[34] The influence of the geometrical and electronic structure, e.g. charges, of the higher indexed step edges should also be considered as an important determining factor.

A third kind of island with a stripe-like shape is also observed in the samples in which type 2 islands are present; two of them are marked with number 3 in Figure 1(c). These islands display a similar apparent height to type 2 pentacene making their classification ambiguous. In order to un-derstand their nature, Kelvin measurements have been analyzed.

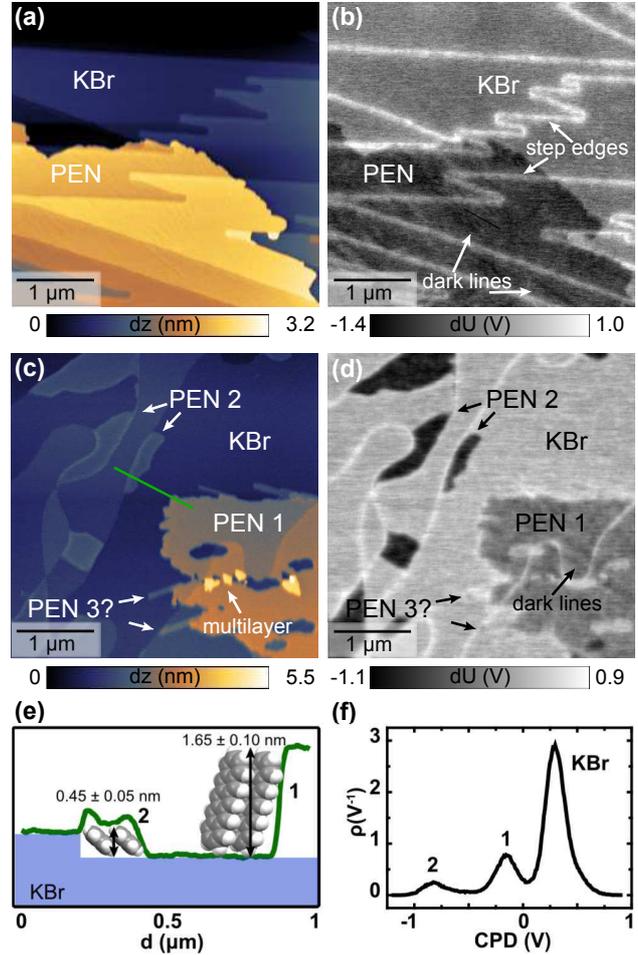

Figure 1: a)-d) Topography and corresponding Kelvin probe images of pentacene islands on KBr(001). a) & b) LCPD differences between the molecular islands, the alkali halide surface, and the substrate step edges are noticeable. $\gamma = 0.08$ fN$\sqrt{m}$. c) & d) Three different types of islands are distinguished. Type 1 is the same kind as the island in image (a & b). Type 2 and 3 display lower step heights. $\gamma = 0.31$ fN$\sqrt{m}$. e) Line profile of the linecut in (c) that determines the topographic height of type 1 and 2 islands. f) Histogram of the LCPD values in (d).

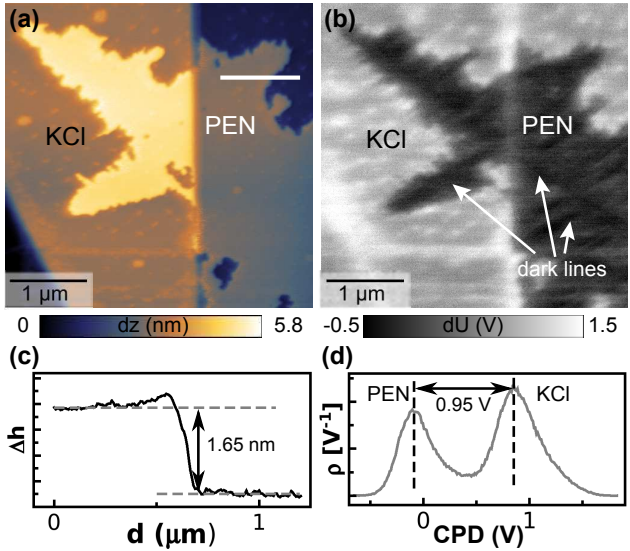

Figure 2: a) & b) Topography and corresponding Kelvin probe image of a pentacene island on KCl(001). $\gamma = 0.24$ fN$\sqrt{m}$. Also on KCl strong LCPD differences between the molecular islands, the alkali halide surface, and the substrate step edges are noticeable. c) Line profile of the linecut in (a). d) Histogram of the LCPD values in (b).

KPFM images in Figure 1(b & d) and Figure 2(b) show that the molecular islands as well as the substrate step edges display a significant LCPD value compared to the terraces. The strong Kelvin contrast at step edges has already been discussed in the literature:[11,35] Monovalent alkali halide type ionic crystals contain a small but noticeable amount of positively charged bivalent impurities such as $Mg^{2+}$ or $Ca^{2+}$. These impurities are distributed evenly in the bulk crystal. The counterparts that maintain the charge neutrality are mostly anion vacancies and gather at step edges at the surface where their coordination is reduced. Thus, the step edges are negatively charged compared to the terraces, which corresponds to positive contrast in KPFM images following the criteria given by Barth et al.[11,13,35,36] On the other hand, type 1 and 2 pentacene islands appear darker than the terraces, meaning either that they are positively charged compared to the substrate, or that a dipole moment is induced within the molecular layer or at the molecule-substrate interface which points upward, i.e., with a positive partial charge at the top of the pentacene layer and a negative partial charge close to the surface.

After analyzing the measured KPFM values we obtain that, on average, upright standing molecular islands on KBr (type 1) have a LCPD with respect to the crystal surface of approximately $-0.45 \pm 0.10$ V, whereas the tilted ones (type 2) show a LCPD of $-1.08 \pm 0.10$ V, see Figure 1(f). On KCl, the LCPD between the upright standing molecules and the substrate is $-0.95 \pm 0.10$ V, see Figure 2(d). The magnitude of these LCPDs is comparable to previous results of the adsorption of molecules on bulk insulators, e.g. PTCDA on NaCl(001)[12] or triphenylene derivatives on KBr(001).[13] Type 3 islands, however, do not show any significant contact potential difference with the substrate. Consequently, we ascribe such stripe-like protrusions to KBr islands produced by slightly overheating during preparation. The KPFM measurements permit us to unambiguously distinguish between differently oriented molecules, since they give raise to different LCPD values. This sensitivity has also been observed for other molecules on metal[4,37] and insulating surfaces.[13]

When the LCPD differs substantially for every compound on the surface of the sample as it occurs in our system, the electrostatic forces need to be properly compensated in order to obtain the correct height.[38] Therefore, the values given here are obtained from measurements performed with the KPFM controller simultaneously with acquiring the topographic images. In the case of polarizable organic materials, in addition, an induced dipole may result from the tip-sample bias which may alter the measured apparent LCPD.[12]

It is important to understand where dipoles or charges are formed in order to find the origin of the different values of the LCPD. It is worth noticing that there is no difference between the signal measured on monolayers compared to multilayers (see Figures 1(c) and 5). This means that the dominant process causing the change in LCPD occurs either at the first molecular layer due to an adsorption-induced polarization, or at the interface between the pentacene and the alkali halide substrate owing to an interface dipole formation. A similar effect has also been observed in the case of thin layers of insulating materials on metals, where the LCPD was independent on the number of insulator layers.[39]

To explore the possible origins of the LCPD density-functional theory (DFT) calculations have been performed. First, the local charge distribution in unperturbed pentacene layers has been considered. As the simplest way of modeling the LCPD values, we have plotted the electrostatic potential on an isodensity surface (see Supporting Information for details). For type 1 islands, mainly the hydrogen atoms are probed by the tip, while for type 2 islands the aromatic $\pi$-system is (at least partially) exposed. In the latter case, a larger negative electrostatic potential due to the $\pi$-electrons should be observed. But this effect is opposite to the one found experimentally. Thus, it can be ruled out the unperturbed local charge distributions in pentacene layers as responsible for the observed LCPD.

From the experimentally obtained LCPD values the interface dipole per molecule can be calculated using the Helmholtz equation:

$$\mu_{mol} = -\varepsilon_0 \Delta U_{CPD} A_{mol} \qquad (1)$$

where $A_{mol}$ is the surface area occupied by one molecule. For type 1 islands we obtain $\mu_{mol} = 9.53 \cdot 10^{-31}$ Cm, assuming that the molecules are completely perpendicular to the surface, and $A_{mol} = a \cdot b/2 = 0.239$ nm$^2$ ($a = 0.790$ nm and $b = 0.606$ nm[24,25]). The tilted molecules (type 2) have a polarization of $\mu_{mol2} = 9.27 \cdot 10^{-30}$ Cm, considering the pentacene as a flat molecule $A_{mol} = c \cdot b = 0.970$ nm$^2$ ($a = 1.601$ nm[24,25]). Thus, the induced dipole moment in one pentacene molecule of type 2 islands is one order of magnitude larger than for the type 1.

We mentioned above that there are divalent impurities in the rocksalts bulk, and that negative charges are located at kink sites in order to compensate these charges.[11] Still, over the terraces this positive charge is not perfectly compensated and an electrostatic field is formed at the sample surface. This electric field can induce an interface dipole moment in the pentacene layer. The direction of the dipole moment induced by a positive background charge is consistent with the LCPD of the pentacene islands with respect to the alkali halide terraces. To estimate the induced dipole moments, the polarizability of a pentacene molecule has been calculated with time-dependent DFT (TDDFT). The polarizability of pentacene is strongly anisotropic and its component perpendicular to the surface is twice as large for molecules in type 1 islands as for those in type 2 islands. Thus, one would expect type 1 islands to show a larger absolute LCPD value than type 2 islands, which is not the case. Hence, an adsorption-induced polarization cannot explain the measured LCPD. Moreover, the adsorption-induced polarization should become larger for multilayers of pentacene, which is not observed experimentally.

This leaves an interface dipole between the molecules and the surface as remaining explanation for the observed LCPD. The apparent positive charge of the pentacene islands compared to the surface is consistent with a partial transfer of electrons from the pentacene molecules to the surface, which is in line with the low ionization potential of pentacene. Using the dipole moments per molecule determined from the LCPD and assuming a separation of 0.5 nm between the partial positive charge in the pentacene molecule and a partial negative countercharge at the surface, this would require the transfer of a partial charge of 0.012 e per pentacene molecule in type 1 islands and of 0.118 e per pentacene molecule in type 2 islands. Considering that in type 2 islands this charge can be distributed over the whole $\pi$-system while in type 1 islands it has to be localized close to the surface, these values appear reasonable. Such a charge transfer at the molecule-surface interface is also consistent with the observation that the LCPD is unchanged for multilayers. Verifying this explanation of the LCPD with DFT calculations would require modeling both the molecular layers and the alkali halide surface, which we did not attempt here.

The stability of the pentacene layer is an important issue for practical applications. It is observed that the pentacene islands adsorbed on the alkali halide (001) surfaces suffer from after-deposition dewetting. Thus, the evolution and changes of the morphology has been investigated on the time scale of hours to days. The post-deposition dewetting of pentacene on KBr and KCl leads to three different visible effects: 1. The substructure of monolayer islands gets more pronounced. 2. The island edges and the area near the edges break and holes appear. 3. Multilayers build up at the expense of monolayers.

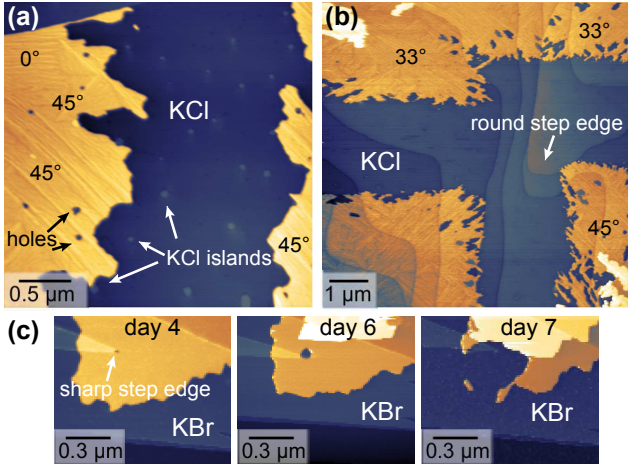

Figure 3: Dewetting of pentacene islands on: a) & b) KCl(001) after of 5 and 8 days, respectively. c) KBr(001) after several days. The borders of the islands fray and holes appear on the terraces, mainly where defects such as step edges are present. a) $\gamma = 0.15$ fN$\sqrt{\text{m}}$. b) $\gamma = 0.09$ fN$\sqrt{\text{m}}$. c) $\gamma = 0.04$ fN$\sqrt{\text{m}}$.

When scanning with lower cantilever amplitudes, a substructure becomes visible on the molecular islands. This substructure gets more pronounced with time, especially for the islands grown on the KCl(001) surface. Figure 3 shows islands several days after deposition. The substructure consists of many line-shaped features organized in some kind of domains with different orientations, see Figure 3(a), (b) and (c). In each domain the substructure lines point to similar directions, which are not randomly oriented but seem to show some preferred angles. For example in Figure 3(a) the substructure lines run mainly along 0°, 45° and 135°, whereas in Figure 3(b), also substructure lines along 33°, 135° and 148° with respect to the (001) axis of KCl are observed. Such orientations correspond to 33°, 45°, and their equivalent directions. Kiyomura et al.[18] experimentally measured with AFM and high-resolution electron diffraction (HRED) that the longer axes of the dendritic layer crystals grown on alkali halide surfaces show preferred epitaxial orientations. On KBr these orientations corresponded to 7° and 33°, and on KCl to 0° and 32° with respect to the [100] direction. In this work,[18] they explained such angles considering the growth as point-on-line epitaxy without misfit along the line. For this epitaxy, a lattice constant of the deposited molecular crystal is modified from the value of its crystalline form to match the length of a basic lattice line of the substrate. Thus, a lattice plane of the molecular crystal is expected to align in parallel with a basic lattice line of the substrate.[18] Kiyomura et al.[18] also found that the measured angles are suitable from the view of lattice misfit, in the following way: For KBr: (110) plane KBr ∥ (11) lattice line of pentacene (pent) for 6.8°, and (200)KBr ∥ (12)pent for 32.4°. For KCl: (100)KCl ∥ (10) pent for 0°, and (200)KCl ∥ (12)pent for 32.4°.[18] Although they also point out that the 0° orientation in KCl could be due to the nucleation at surface steps along the [100] direction.[19] Taking into account these observations, we can relate the substructure of the islands to the epitaxial growth of the pentacene islands. We believe that it gets more prominent with time due to a change in the arrangement of the molecules by dewetting for forming more stable crystallites.

The morphology of the island edges is also affected by dewetting: the borders fray and holes break, as it can be seen in the images of Figure 3. The holes do not break randomly but their position is correlated with the position of surface defects. Small differences in the preparation conditions of the alkali halide surfaces have as consequence the appearance of slightly different surface defects. Figure 3 illustrates the role of such surface defects. In Figures 3(a) and (b) round step edges and small KCl islands are visible. Here, the holes break preferably above the small islands. The sample in Figure 3(c) was annealed at slightly lower temperatures and shows sharp step edges and clean terraces. On this surface, holes only appear at the sharp step edges. In Figure 3(b), the initial borders of the molecular islands previous to dewetting are still recognizable. The islands displayed a squared shape aligned with the <100> directions of the KCl substrate. The holes typically elongate along the directions that are followed by the substructure of the pentacene islands. With the formation of such holes, the island's shape modifies into a fractal-like geometry. The change of the pentacene island morphology towards dentritic structures has

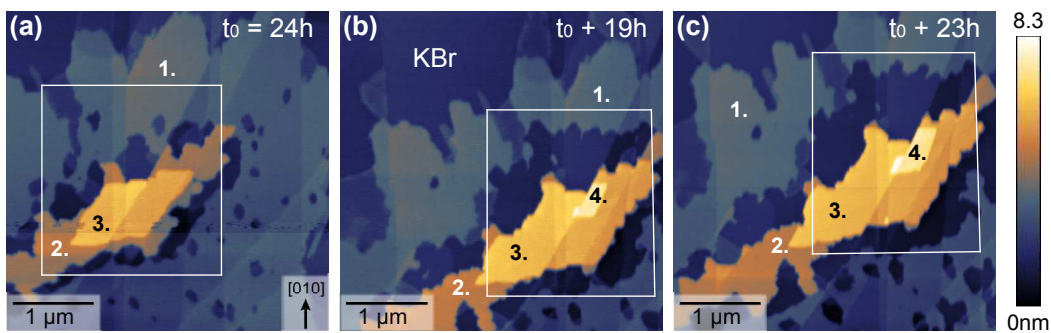

Figure 4: Temporal evolution of the topography of pentacene islands on KBr(001). The multilayer island at the center increases in height and area, while an increasingly large trench is formed between the multilayer and the adjacent monolayer islands. a) $\gamma = 0.31$ fN$\sqrt{m}$. b) $\gamma = 0.05$ fN$\sqrt{m}$. c) $\gamma = 0.06$ fN$\sqrt{m}$.

also been observed on SiO$_2$ in thermally activated dewetting experiments.[40]

Figure 4 displays the temporal evolution of a pentacene island grown on KBr. In the center of Figure 4(a), there is a multilayer island which contains a second and a third layer (marked with numbers in the images). The multilayer island is separated by a trench from the surrounded monolayer islands. In Figure 4(b), the multilayer island has changed: a fourth layer has appeared. This new layer gradually increases its surface with time (see Figure 4(c)). Simultaneously to this growth, the surrounding monolayer islands decrease in size and the trench grows. The same effect is visible in the sequence of Figure 3(c) where a multilayer grows while the hole in the lower layer increases in size. The disappearance of monolayers to form higher-layered islands is a common way of dewetting of molecules on alkali halide surfaces.[41] We expect the new layers to grow directly in the bulk phase configuration, since it is more stable than the thin film phase.[27,28] This new arrangement may also explain the increase of visibility of the substructures, since the molecules have a larger tilting angle in the bulk phase than in the thin film phase.[27,28]

In order to better understand the dewetting of pentacene on alkali halide surfaces further analyses of the Kelvin probe images have been performed. Figure 5 presents topographic and KPFM data of dewetted molecular islands on KBr and KCl. In Figure 5(b), type 1 and type 2 islands of the differently oriented molecules reported above are clearly distinguishable. We mentioned above that a substructure in the topographic images of type 1 islands was unveiled. In the KPFM images, also a substructure is observed, it consists of dark lines that cross the type 1 islands, arrows in Figure 1(b & d) and Figure 2(b). The LCPD value at such dark lines is larger than the one of type 1 islands but lower than the type 2 one. Since the difference in LCPD originates from the difference in molecular orientation, the relation of these dark lines to a rearrangement of pentacene molecules is supported.

When the islands are affected by dewetting, their morphology changes and multilayers grow at the expense of monolayer islands. As previously discussed, these multilayers show the same surface potential as the monolayer islands. However, the Kelvin signal of such multilayer islands displays an additional feature that is not correlated with any visible topographic change, or at least not along all layers of the islands: a dark line runs along its middle axis. Figure 5(b & d) show multilayer islands grown on KBr and KCl, respectively, at these images dark lines crossing the newly formed islands are clearly seen. Such dark lines are also observed in dewetted monolayer islands, such as the one in Figure 2, where also secondary dark lines are seen. However, this dark line in multilayer islands is predominant. For the sake of clarifying the role of this feature, a thorough analysis of the position and the orientation of the dark lines in the acquired data has been done, in particular, in the dewetted multilayer islands. Taken into account a margin of error owing to drift in our scans, the results show that: first, when the lines are visible on multilayers they always run along the middle axis of the islands. Second, the dark lines in KPFM follow preferred orientations. The angles of such orientations with respect to the [100] correspond to 33° and 45° for pentacene islands on KCl, and 0°, 33° and 45° on KBr, and their equivalent directions. These angles match the orientation of the substructure lines also observed in the topography. We can therefore, as we did with the topographic substructure, relate the dark lines of the KPFM to the epitaxial formation of the pentacene crystals.

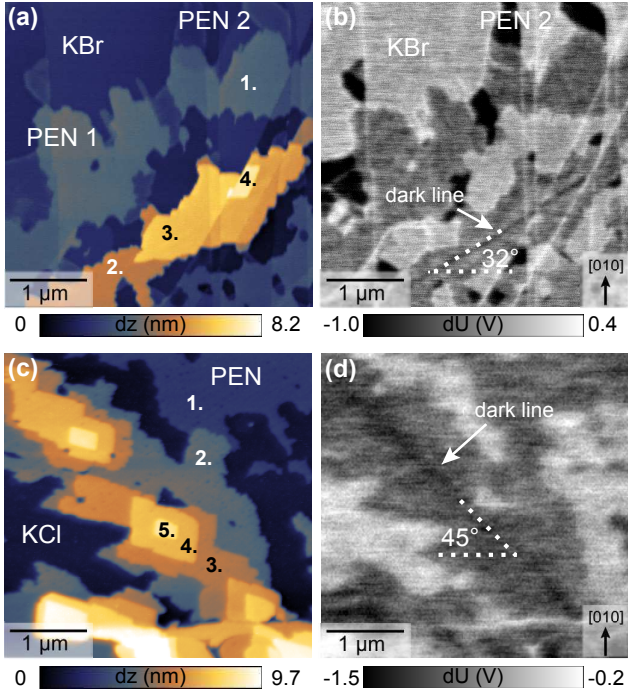

Figure 5: Topography and corresponding Kelvin probe image of the dewetting of pentacene islands on: a) & b) KBr(001) after 2 days. $\gamma = 0.05$ fN$\sqrt{\text{m}}$. c) & d) KCl(001) after 4 days. $\gamma = 0.56$ fN$\sqrt{\text{m}}$. Surprisingly there are no differences in the LCPD of multilayers and monolayers. A dark line at the center of the multilayer island is revealed in the Kelvin probe images.

Summarizing, after deposition, the pentacene organize in monolayer islands. Induced by dewetting, the molecules start to rearrange towards a lower energetic configuration. This conformational change is accompanied by modifications of the islands topography, including the formation of multilayer islands, while the surrounding monolayer islands disappear. The newly formed dewetted multilayer islands display a favored alignment, i.e. we are able to follow the growth of the pentacene crystals from their initial stage. The preferred directions in our system 0° and 45° may correspond to nucleation controlled at step edges in which the growth direction is given by the orientation of the step edge. On the other hand, the 33° is given by the point-on-line epitaxy of the pentacene molecular crystal with respect to the alkali halide surface, confirming the observation of Kiyomura et al.[18,19]. The appearance of the dark lines in the surface potential is however not straightforward to explain. We have discussed throughout this paper that the Kelvin contrast arises from the interface alkali halide-molecules. The pentacene molecules have a weak epitaxial growth on KBr and KCl (001) surfaces, namely only a point-of-line epitaxy. Therefore we expect the interaction molecule-substrate not to be homogeneous over the whole surface. At the starting growing line of the crystal, i.e. the axis of the point-on-line epitaxy, this interaction should be maximal giving rise to the strong KPFM signal. The interaction can be electrostatic (e.g., changes in the interface dipole or charge accumulation), or mechanical, such as strain. Dewetting is known to be driven by strain in the molecular layer, which may be released by changing its internal structure.[7]

Type 2 molecular islands behave in a completely different way than the type 1 ones. Their borders do not change much, their topography and Kelvin contrast do not vary, neither do they form multilayer. Some of the small islands surrounded by type 1 islands disappear or are "absorbed" by the growth of multilayer, as we can see in Figure 4. On the other hand, the isolated ones enlarge their size along the high-indexed step edges at which they grow and remain more ore less stable in time, as it is shown in Figure 6.

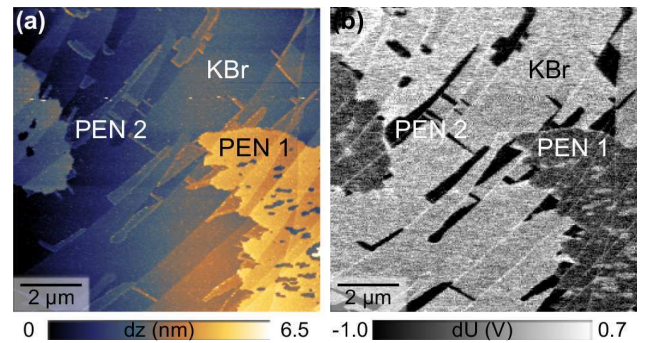

Figure 6: a) & b) Topography and corresponding Kelvin probe image of the dewetting of pentacene type 2 islands on KBr(001) after 3 days. $\gamma = 0.08$ fN$\sqrt{\text{m}}$.

# Conclusions

We have studied the growth and electronic properties of pentacene on KBr and KCl (001) surfaces, and found that after deposition mainly islands of upright standing molecules are formed. The islands show a strong LCPD compared to the one of the substrate. After comparing our experimental results with DFT calculations, we explain this LCPD as the result of an interface dipole caused by a partial transfer of electrons from the molecules to the alkali halide surface. The morphology of the islands is strongly influenced by post-deposition dewetting: metastable monolayer pentacene islands evolve within days of deposition into more stable multilayer crystals which show preferential orientations. The new multilayer islands display a new feature in the KPFM images, a dark line which runs along their middle axis. We relate such lines to the epitaxial growth of pentacene crystals on the alkali halide surfaces, their dark contrast is probably produced by accumulated charge at domain boundaries. Summarizing, we have shown that the crystallization process of molecules can be monitored by NC-AFM and KPFM experiments.

# Methods

**Experimental Details** The sample preparation and measurements were carried out in an ultra-high vacuum chamber with a base pressure of less than $3 \cdot 10^{-10}$ mbar. Atomically clean KBr and KCl (001) surfaces were obtained by cleaving single crystals in air, immediately introducing the crystals to the vacuum chamber and heating them to 400 K for 1 h. The pentacene molecules were thermally deposited onto the substrates after degassing the molecular source for several hours at temperatures slightly below the sublimation temperature (508 K). Samples were then transferred to the Omicron scanning force microscope (Omicron NanoTechnology GmbH, Germany) equipped with Nanosensors cantilevers (Neuchatel, Switzerland) and a Nanonis Phase-locked loop electronic (SPECS, Switzerland). All measurements were carried out in the non-contact mode, where the tip is oscillated at an amplitude of a few nm kept constant by a feed-back loop at resonance. The resonance frequency of the cantilever is measured as it reduces when the tip is approached to the sample surface under the influence of the interaction of tip and surface. Topographical imaging is carried out at constant frequency shift using cantilevers with a force constant of 40 N/m and a free resonance frequency of 170 kHz. For Kelvin probe measurements performed in parallel to the topography measurements, the voltage applied to the tip was oscillated with a frequency of 2 kHz and an amplitude of 3 V (frequency-modulation-mode of KPFM). For these measurements, cantilevers with a Platin-Iridium coated tip, a force constant of 3 N/m, and a free resonance frequency of 75 kHz were used. For characterizing the NC-AFM images the normalized frequency shift has been used: $\gamma = \Delta f k A^{3/2}/f_0$.

**Computational Details** DFT calculations were performed with the Amsterdam Density Functional (ADF) program package.[42,43] The BLYP exchange–correlation functional[44,45] was used in conjunction with the TZ2P Slater-type basis set.[46]

**Acknowledgement** We thank the European Research Council for financial support through the starting grant NANOCONTACTS (No. ERC 2009-Stg 239838), and Michael Marz for fruitful discussions.

*Supporting Information Available:* Electrostatic potential on an isodensity surface for a cluster of three pentacene molecules.

# Notes and References

# Supporting information for:

# Epitaxial Growth of Pentacene on Alkali Halide Surfaces Studied by Kelvin Probe Force Microscopy

Julia L. Neff[†], Peter Milde[‡], Carmen Pérez León[†], Matthew D. Kundrat[§],

Christoph R. Jacob[§], Lukas Eng[‡], Regina Hoffmann-Vogel[†*]

[†]Physikalisches Institut and DFG-Center for Functional Nanostructures,

Karlsruhe Institute of Technology, Wolfgang-Gaede-Str. 1, 76131 Karlsruhe, Germany

[‡]Institut für Angewandte Photophysik, Technical University of Dresden,

George-Baehr-Strasse 1, 01069 Dresden, Germany

[§]Institute of Physical Chemistry and DFG-Center for Functional Nanostructures,

Karlsruhe Institute of Technology, Wolfgang-Gaede-Str. 1a, 76131 Karlsruhe, Germany[*]

E-mail: r.hoffmann@kit.edu

---

[*]To whom correspondence should be addressed

To access whether the measured LCPD between type 1 and type 2 islands can be caused by the local charge distribution in unperturbed pentacene, we performed DFT calculations for a pentacene molecule and a small cluster of three pentacene molecules extracted from the bulk crystal structure.[1] Note that for such a cluster of three molecules, an almost identical structure can be extracted from the cyrstal structure of the thin-film phase. For this cluster, the electrostatic potential on an isodensity surface is shown in Fig. 1. On the left, a top view of type 2 islands is shown, while on the right a top view of type 1 islands is plotted. In the first case, the average electrostatic potential is approximately 0.05 atomic units, 0.05 a.u., (= 1.4 V) more negative than in the second case. This difference depends on the chosen isodensity value and decreases with decreasing isodensity values (i.e., for a larger distance of the tip from the pentacene layer). Nevertheless, the qualitative picture is independent of the chosen isodensity value.

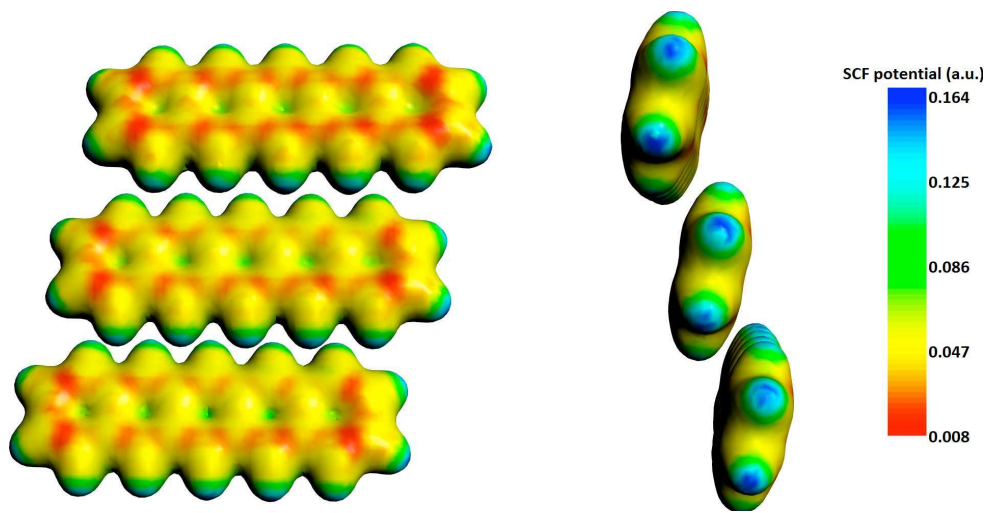

Figure S1: Electrostatic potential on an isodensity surface (corresponding to a density of 0.015 e/bohr$^3$) for a cluster of three pentacene molecules. Top view images of the type 1 islands, on the right, and type 2 islands, on the left.

For evaluating the extent to which the adsorption-induced polarization can account for the observed LCPD, we calculated the polarizability of a single pentacene molecule with TDDFT using the geometry extracted from the crystal structure. We find polarizability components of $\alpha_{xx} = 216$ a.u., $\alpha_{yy} = 768$ a.u., and $\alpha_{zz} = 1365$ a.u., where the x axis is perpendicular to the molecular plane, the y axis is in the molecular plane pointing in the short direction of the molecule and the z axis coincides with the molecular axis. Using a tilting angle of 70°, this corresponds to a polarizability component perpendicular to the surface of 1357 a.u. for a pentacene molecule in a type 1 island. Similarly, using a tilting angle of 30° for type 2 islands, we find a polarizability component perpendicular to the surface of 571 a.u. Thus, for type 1 islands an induced dipole moment approximately twice as large as for type 1 islands is expected.